\documentclass[journal]{IEEEtran}

\usepackage{graphicx}
\usepackage{cite}
\ifCLASSINFOpdf
\else
\fi
\usepackage[cmex10]{amsmath}
\usepackage{algorithmic}
\usepackage{url}
\usepackage{algorithm}
\usepackage{algorithmic}
\usepackage{subcaption}
\usepackage{float}

\begin{document}

\title{Grid-Connected Emergency Back-Up Power Supply}

\author{Dhiman Chowdhury,~\IEEEmembership{Student Member,~IEEE}, Mohammad Sharif Miah, Md. Feroz Hossain,  Md. Mostafijur Rahman, Md. Marzan Hossain, Md. Nazim Uddin Sheikh, Md. Mehedi Hasan, Uzzal Sarker and Abu Shahir Md. Khalid Hasan
\thanks{Dhiman Chowdhury is with the Department
of Electrical Engineering, University of South Carolina, Columbia,
SC 29208, U.S.A. (e-mail: dhiman@email.sc.edu)}
\thanks{Mohammad Sharif Miah, Md. Feroz Hossain,  Md. Mostafijur Rahman, Md. Marzan Hossain, Md. Nazim Uddin Sheikh, Md. Mehedi Hasan and Uzzal Sarker are with the Department of Electrical and Electronics Engineering, Daffodil International University, Dhaka 1207, Bangladesh.}%
\thanks{Abu Shahir Md. Khalid Hasan is with the Department of Electrical and Electronic Engineering, Bangladesh University of Engineering and Technology, Dhaka 1000, Bangladesh.}} 
%
%
%
%
\markboth{Submitted to arXiv.org}%
{Shell \MakeLowercase{\textit{et al.}}: Bare Demo of IEEEtran.cls for IEEE Journals}

\maketitle

\begin{abstract}
This paper documents a design and modelling of a grid-connected emergency back-up power supply for medium power applications. There are a rectifier-link boost derived battery charging circuit and a 4-switch push-pull power inverter circuit which are controlled by pulse width modulation (PWM) signals. This paper presents a state averaging model and Laplace domain transfer function of the charging circuit and a switching converter model of the power inverter circuit. A changeover relay based transfer switch controls the power flow towards the utility loads. During off-grid situations, loads are fed power by the proposed inverter circuit and during on-grid situations, battery is charged by an ac-link rectifier-fed boost converter. There is a relay switching circuit to control the charging phenomenon of the battery. The proposed design has been simulated in PLECS and the simulation results corroborate the reliability of the presented framework.
\end{abstract}

\begin{IEEEkeywords}
Back-up power supply, Laplace domain, push-pull inverter, state averaging model, switching converter model
\end{IEEEkeywords}

\IEEEpeerreviewmaketitle

\section{Introduction}
\IEEEPARstart{P}{ower} electronic converters can provide alternate and sustainable solutions to grid power failure, poor voltage regulation and inefficient and expensive mains power supply systems. In the events of grid power unavailability, emergency utility loads and critical loads like computer and different medical devices can be supplied power by power electronic devices like instant power supply (IPS) and uninterruptible power supply (UPS). In both of IPS and UPS systems, power inverter circuits are designed to efficaciously deliver power to the utility loads. The basic difference between IPS and UPS is the back-up power switching transfer time.

In this paper a medium power grid-connected switching converters based emergency back-up power supply has been presented which can be used as a reliable UPS or IPS or both the systems. There is a 4-switch push-pull inverter circuit which supports the utility loads when the mains power is unavailable. The battery is charged by a rectifier-link dc-dc boost converter circuit. There is an electrical isolation at the source-end of the battery charger which steps-down the line voltage and the converter produces a suitable voltage level to charge the battery. The switching phenomenon of the boost converter has been controlled by high frequency PWM signals to reduce the current ripples, size of the filter components and switch conduction losses. The converter has been designed in such a manner that it operates in continuous conduction mode (CCM) so that the steady state inductor current is greater than the ripple components.

A 4-switch push-pull inverter circuit yields to large current driving capabilities to deliver sufficient power to the loads. There are parallel snubber components with the inverter switches to reduce $\frac{dv}{dt}$ effects. There is a 50 Hz center-tapped step-up transformer which is followed by an L-C low pass filter at the load side. The inverter switches are controlled by two complementary PWM signals of the mains line frequency.

During on-grid condition, the loads and the battery are fed power by the mains line and the grid-tied charger respectively. During off-grid condition, the customized power supply system delivers power to the loads. The power transfer switching from grid to the customized power supply system is automatic and instantaneous. The transfer application has been substantiated by a changeover relay with a switch operating rate of 3-5 ms. There is a switching relay circuit to control the charging process of the battery. If the battery voltage is at its rated value, the charge controller disconnects the battery from the charger and thus prevents the over-charging phenomenon.

\begin{figure*}
	\centering
	\includegraphics[height=5.0in, width=7.0in]{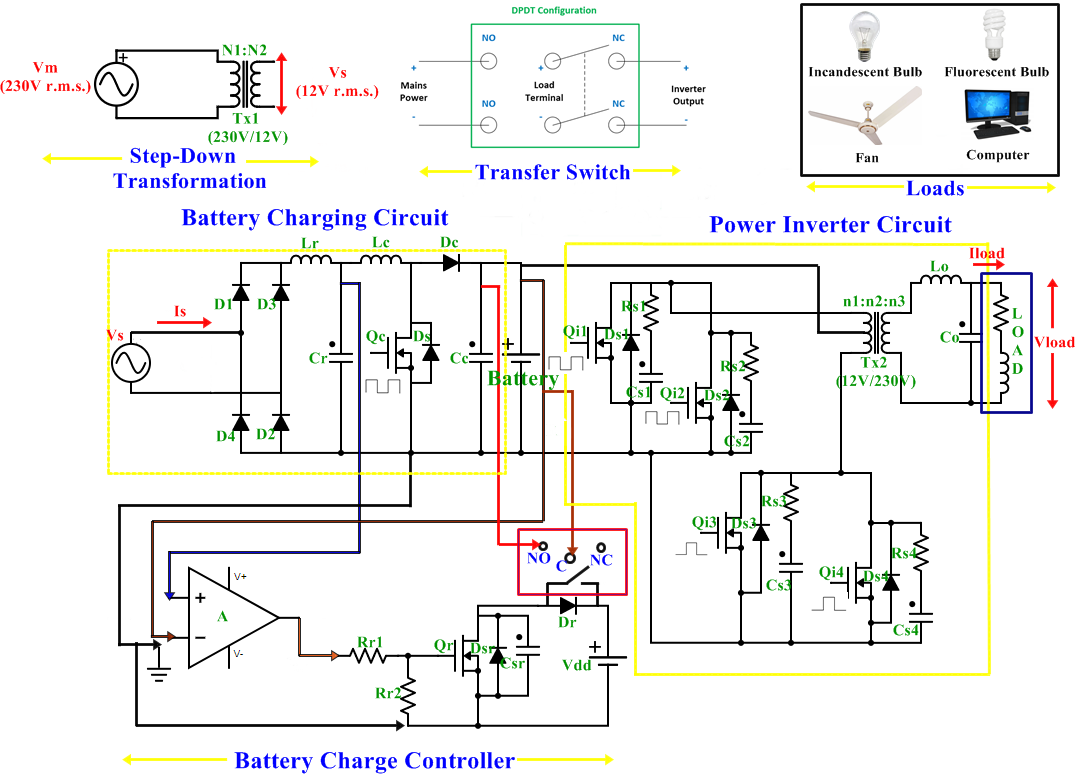}
	\caption{Proposed system layout}
	\label{fig1}
\end{figure*}
For effective and sustainable power supply solutions, many novel PWM inverter topologies have been proposed and reported in \cite{1} - \cite{8}. A 250 w soft-switching current-fed push-pull converter topology has been presented in \cite{9}. For a single-phase PWM inverter, a current quality evaluation methodology has been presented in \cite{10}. For performance betterment, reliable PWM control techniques for inverter circuits have been proposed in \cite{11} and \cite{12}. The proposed push-pull inverter based emergency power supply system can be used as a UPS device to support critical loads. Several inverter topologies and modelling approaches for UPS applications have been reported in \cite{13} - \cite{16}. Immaculate parallel UPS operations and control techniques of single phase inverters for UPS systems have been articulated in \cite{17} - \cite{21}.

In this paper, a state averaging model of the battery charging circuit has been derived and a Laplace domain transfer function has been determined from the time domain model. The inverter circuit has been analyzed in the form of an equivalent switching converter model. The proposed system has been simulated in PLECS for an R-L load. Harmonic analysis of mains current, inverter output voltage and load current have been conducted to evaluate the corresponding total harmonic distortion (THD) values. The sending-end power factor has been calculated as well to verify the reliability of the proposed system. Sec. II subsumes the description of the proposed design, sec. III contains the state space model of the battery charging circuit, sec. IV presents the switching converter model of the power inverter, sec. V documents the simulation results and sec. VI concludes the article. 

%
%
\section{Proposed Design}
A grid-integrated emergency back-up power supply with automatic transfer switching application between mains power and customized power inverter circuit is presented in this paper. A 4-switch push-pull inverter with snubber components and an iron-core step-up output transformer has been designed and implemented to feed loads in the absence of grid power. A rectifier-fed boost converter has been developed to charge the battery at a suitable voltage level. Fig. \ref{fig1} presents the proposed system layout.

The transfer switching operation from the mains line to the customized power circuit in case of grid power failure has been implemented by a changeover relay following a double pole double throw (DPDT) switching structure. Direct connection relay $CR_{1}$ is the relay terminal connected to the mains power and it bridges the connection between grid line and utility load. Circuit-to-load connection relay $CR_{2}$ is the relay terminal connected to the inverter output and it bridges the connection between the customized power supply unit and utility load. $CR_{1}$ and $CR_{2}$ get activated alternatively and in general case, $CR_{2}$ operates when mains power is unavailable. A generic relay has two switching terminals and one moving pole to switch from one terminal to another. In this DPDT relay, the normally closed ($NC$) terminal or $CR_{2}$ is connected to the inverter output and the normally open ($NO$) or $CR_{1}$ is connected to the mains line. The moving pole is connected to the load. In the de-energized state, which means mains power is off, the load is connected to $NC$ and conversely, in the energized state, which means mains power is available, the load gets automatically connected to $NO$.

An intelligible relay switching circuit has been
designed which determines the connectivity of the battery to
the charger. If the battery voltage is at its rated nominal value, the switching circuit disconnects the charger from the battery to provide over-voltage and over-heat protection. Charger connection relay $CCR$ is basically a single pole double throw (SPDT) relay switching circuit to connect the battery terminal to the charger. Battery charge controller consists of a comparator circuit in which the reference voltage $V_{ref}$ is fed from the rectifier output and it is connected to a non-inverted port and the battery voltage $V_{bat}$ is connected to an inverted port of an op-amp. The difference voltage $V_{diff}$ is [$V_{ref}-V_{bat}$] and the comparator output voltage $V_{comp}$ is $-V_{sat} = V-$, if $V_{bat}\geq V_{ref}$ and is $+V_{sat} = V+ $, if $V_{bat} < V_{ref}$; here $V_{sat}$ is the saturation voltage. The comparator circuit is followed by a relay switching circuit of which the input is the comparator output voltage. Here $NO$ terminal is connected to the charger, $NC$ terminal is open and the moving pole, $C$ is connected to the battery. When $V_{comp} = V-$, there is no current flowing through the switching relay and the battery is disconnected from the charger terminal. When $V_{comp} = V+$, a current flows through the relay to close the switch and the battery is connected to the charger.

\begin{figure}[!t]
	\centering
	\includegraphics[height=4.3in ,width=3.5 in]{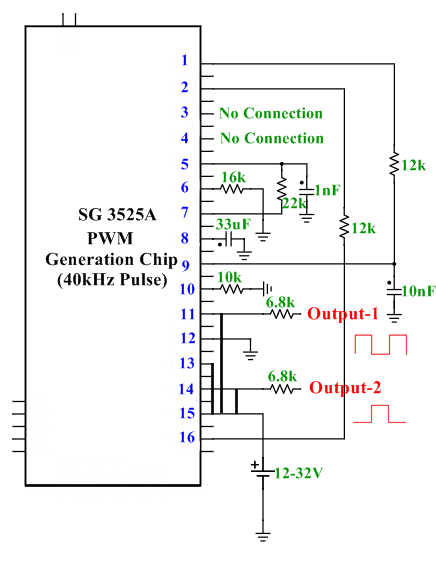}
	\caption{40 kHz PWM signal generation circuit using SG3525A}
	\label{fig2}
\end{figure}

\begin{figure}[!t]
	\centering
	\includegraphics[height=4.3in ,width=3.5 in]{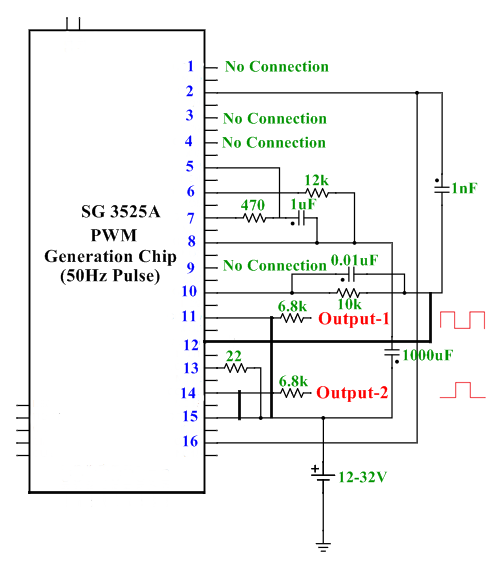}
	\caption{50 Hz PWM signal generation circuit using SG3525A}
	\label{fig3}
\end{figure}

The switching operations of the boost converter and push-pull inverter can be executed by PWM signals of 40 kHz and 50 Hz respectively. Analog integrated chip (IC) SG3525A can be used to generate the switch-control PWM signals. 
SG3525A generally produces two complementary pulses of the same frequency. In case of the boost converter, any of the two generated 40 kHz pulses can be used to control the switching operation. In regard to maintain the trade-off between switching loss and conduction loss of a switching device, 40 kHz switching frequency has been optimized here. In case of the push-pull inverter, both of the 50 Hz pulses have been used to control the switching operations of the inverter legs, leg1: $Q_{i1}-Q_{i2}$ and leg2: $Q_{i3}-Q_{i4}$. Fig. \ref{fig2} and Fig. \ref{fig3} present the designs of 40 kHz and 50 Hz PWM signal generation circuits respectively.

\section{State Averaging Model of the Proposed Battery Charging Circuit}
\begin{figure}[!t]
	\centering
	\includegraphics[height=2in ,width=3.5 in]{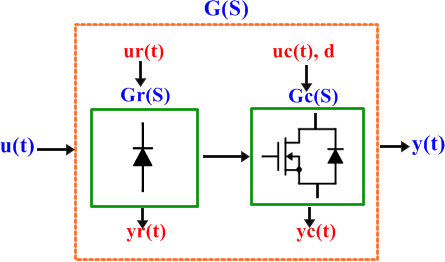}
	\caption{State space structure and transfer function model of the battery charging circuit (two series blocks, $G_{r}(S)$ and $G_{c}(S)$, denote s-domain transfer functions of the rectifier and boost converter respectively)}
	\label{fig5}
\end{figure}
\begin{figure}[!t]
	\centering
	\includegraphics[height=6in ,width=3.2 in]{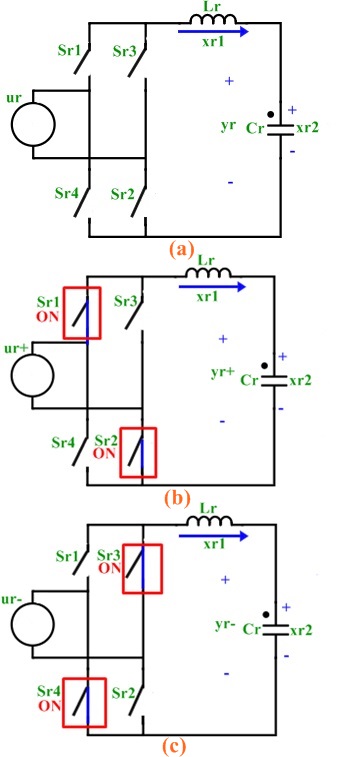}
	\caption{State averaging model of the rectifier circuit; (a) state space model, (b) equivalent model for positive input and (c) equivalent model for negative input}
	\label{fig6}
\end{figure}
\begin{figure}[!t]
	\centering
	\includegraphics[height=4.5in ,width=3.2 in]{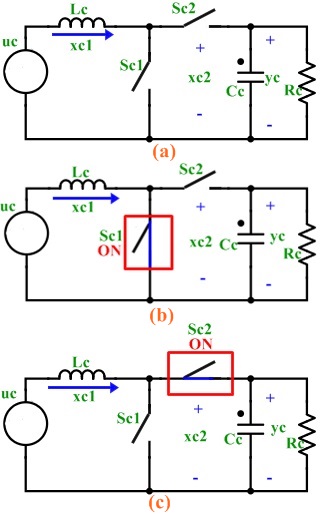}
	\caption{State averaging model of the boost converter; (a) state space model, (b) equivalent model during subinterval-1 and (c) equivalent model during subinterval-2}
	\label{fig7}
\end{figure}
The state space averaging technique implies an approximation methodology to analyze switching converters being operated with high switching frequencies. The state space modelling yields to a continuous-time signal frequency analysis and results in a non-linear system though the original circuit is a linear system. With small signal approximations, state space modelling provides good stability and time variant behavioral analysis of switching converters. The generic state space model equations are
\begin{subequations}
	\begin{align}
		\vec{\dot{x}}(t)=A\vec{x}(t)+B\vec{u}(t)\\
		\vec{y}(t)=C\vec{x}(t)+D\vec{u}(t)
	\end{align}
\end{subequations}
Here $\vec{x}(t)$ is the vector of state variables to be analyzed (inductor current and capacitor voltage),  $\vec{\dot{x}}(t)=\frac{d\vec{x}(t)}{dt}$, $\vec{u}(t)$ is the input source vector (voltage or current source), $\vec{y}(t)$ is the output vector (output voltage) and A, B, C and D are constant matrices. Laplace transformations of the state equations yield to transfer function of a circuit system which is a good determinant of the input-output relationship. Corresponding Laplace domain transformations of (1a) and (1b) to get the system transfer function are
\begin{subequations}
	\begin{align}
		S\vec{X(S)}=A\vec{X(S)}+B\vec{U(S)}\\
		\vec{X(S)}=[(SI-A)^{-1}B]\vec{U(S)}
	\end{align}
\end{subequations}
\begin{subequations}
	\begin{align}
		\vec{Y(S)}=C\vec{X(S)}+D\vec{U(S)}\\
		\vec{Y(S)}=[C(SI-A)^{-1}B+D]\vec{U(S)}\\
		G(S)=\frac{\vec{Y(S)}}{\vec{U(S)}}=[C(SI-A)^{-1}B+D]
	\end{align}
\end{subequations}
Here $I$ is an m$\times$n identity matrix, $\vec{X(S)}$, $\vec{U(S)}$ and $\vec{Y(S)}$ are the s-domain transformations of $\vec{x}(t)$, $\vec{u}(t)$ and $\vec{y}(t)$ respectively. $G(S)$ is the system transfer function. Fig. \ref{fig5} presents the overall state space configuration and transfer function model of the battery charger in which the system transfer function can be expressed as
\begin{equation}
G(S)=G_{r}(S)G_{c}(S)
\end{equation}
here $G_{r}(S)$ and $G_{c}(S)$ are the transfer functions of the rectifier and boost converter circuits respectively. To reduce redundancy of terminology, any time domain variable, $m(t)$, is denoted as $m$ in further expressions in this paper. 
\subsection{Derivation of $G_{r}(S)$}
Fig. \ref{fig6} presents the equivalent state space model of the rectifier circuit where $u_{r}$ is the input ac voltage (stepped-down), $y_{r}$ is the output voltage and $x_{r1}$ and $x_{r2}$ are the state variables. $S_{r1}$, $S_{r2}$, $S_{r3}$ and $S_{r4}$ are the state space models of $D_{1}$, $D_{2}$, $D_{3}$ and $D_{4}$ respectively. The rectifier circuit can be analyzed separately during two half (positive and negative) cycles of the input source such that
\begin{equation}
u_{r}=u_{r+}+u_{r-}
\end{equation}
During positive half cycle of the input
\begin{equation}
\dot{x_{r1}}=-\frac{1}{L_{r}}x_{r2}+\frac{1}{L_{r}}u_{r+}, \dot{x_{r2}}=\frac{1}{C_{r}}x_{r1}, y_{r+}=x_{r2} 
\end{equation}
\begin{subequations}
	\begin{align}
	A_{r1}=
	\begin{bmatrix}
	0 & -\frac{1}{L_{r}}  \\
	\frac{1}{C_{r}} & 0
	\end{bmatrix}
	\\
	B_{r1}=
	\begin{bmatrix}
	\frac{1}{L_{r}}  \\
	0
	\end{bmatrix}, 
	C_{r1}=
	\begin{bmatrix}
	0 & 1
	\end{bmatrix},
	D_{r1}=0
	\end{align}
\end{subequations}
From (3c), (7a) and (7b), the rectifier transfer function during +ve half cycle of the input is
\begin{equation}
G_{r1}(S)=\frac{1}{1+S^{2}L_{r}C_{r}}
\end{equation}
In this follow-up, during negative half cycle of the input
\begin{equation}
\dot{x_{r1}}=-\frac{1}{L_{r}}x_{r2}-\frac{1}{L_{r}}u_{r-}, \dot{x_{r2}}=\frac{1}{C_{r}}x_{r1}, y_{r-}=x_{r2} 
\end{equation}
\begin{subequations}
	\begin{align}
	A_{r2}=
	\begin{bmatrix}
	0 & -\frac{1}{L_{r}}  \\
	\frac{1}{C_{r}} & 0
	\end{bmatrix}
	\\
	B_{r2}=
	\begin{bmatrix}
	-\frac{1}{L_{r}}  \\
	0
	\end{bmatrix}, 
	C_{r2}=
	\begin{bmatrix}
	0 & 1
	\end{bmatrix},
	D_{r2}=0
	\end{align}
\end{subequations}
From (3c), (10a) and (10b), the rectifier transfer function during negative half cycle of the input is
\begin{equation}
G_{r2}(S)=-\frac{1}{1+S^{2}L_{r}C_{r}}
\end{equation}
From circuit analysis and in accordance with (8) and (11), the resultant rectifier transfer function can be expressed as
\begin{equation}
G_{r}(S)=||G_{r1}(S)||=||G_{r2}(S)||=\frac{1}{1+S^{2}L_{r}C_{r}}
\end{equation}
\subsection{Derivation of $G_{c}(S)$}
Fig. \ref{fig7} presents the equivalent state space model of the boost converter circuit where $u_{c}$ is the input dc voltage (rectified+filtered), $y_{c}$ is the output voltage and $x_{c1}$ and $x_{c2}$ are the state variables. $S_{c1}$ and $S_{c2}$ are the state space models of $Q_{c}$ and $D_{c}$ respectively. During subinterval-1 or $dT_{p}$ interval, where d is the duty
ratio and $T_{p}$ is the switching period, $S_{c1}$ is on (closed switch) and during subinterval-2 or $(1-d)T_{p}$ interval, $S_{c2}$ is on (closed switch). During $dT_{p}$ interval,
\begin{equation}
\dot{x_{c1}}=\frac{1}{L_{c}}u_{c}, \dot{x_{c2}}=-\frac{1}{R_{c}C_{c}}x_{c2}, y_{c}=x_{c2}
\end{equation} 
\begin{subequations}
	\begin{align}
	A_{c1}=
	\begin{bmatrix}
	0 & 0  \\
	0 & -\frac{1}{R_{c}C_{c}}
	\end{bmatrix}
	\\
	B_{c1}=
	\begin{bmatrix}
	\frac{1}{L_{c}}  \\
	0
	\end{bmatrix}, 
	C_{c1}=
	\begin{bmatrix}
	0 & 1
	\end{bmatrix},
	D_{c1}=0
	\end{align}
\end{subequations}
During $(1-d)T_{p}$ interval,
\begin{equation}
\dot{x_{c1}}=-\frac{1}{L_{c}}x_{c2}+\frac{1}{L_{c}}u_{c}, \dot{x_{c2}}=\frac{1}{C_{c}}x_{c1}-\frac{1}{R_{c}C_{c}}x_{c2}, y_{c}=x_{c2} 
\end{equation}
\begin{subequations}
	\begin{align}
	A_{c2}=
	\begin{bmatrix}
	0 & -\frac{1}{L_{c}} \\
	\frac{1}{C_{c}} & -\frac{1}{R_{c}C_{c}}
	\end{bmatrix}
	\\
	B_{c2}=
	\begin{bmatrix}
	\frac{1}{L_{c}}  \\
	0
	\end{bmatrix}, 
	C_{c2}=
	\begin{bmatrix}
	0 & 1
	\end{bmatrix},
	D_{c2}=0
	\end{align}
\end{subequations}
In cumulative form,
\begin{subequations}
	\begin{align}
	A_{c}=dA_{c1}+(1-d)A_{c2}\\
	B_{c}=dB_{c1}+(1-d)B_{c2}
	\end{align}
\end{subequations}
Therefore
\begin{subequations}
	\begin{align}
	A_{c}=
	\begin{bmatrix}
	0 & -\frac{1-d}{L_{c}} \\
	\frac{1-d}{C_{c}} & -\frac{1}{R_{c}C_{c}}
	\end{bmatrix}
	\\
	B_{c}=
	\begin{bmatrix}
	\frac{1}{L_{c}}  \\
	0
	\end{bmatrix}, 
	C_{c}=
	\begin{bmatrix}
	0 & 1
	\end{bmatrix},
	D_{c}=0
	\end{align}
\end{subequations}
From (3c), (17a) and (17b), the converter transfer function can be expressed as
\begin{equation}
	G_{c}(S)=\frac{1-d}{S^{2}L_{c}C_{c}+\frac{SL_{c}}{R_{c}}+(1-d)^{2}}
\end{equation}
\subsection{Derivation of $G(S)$}
According to (4), (12) and (19), the overall charging system transfer function can be expressed as
\begin{equation}
G(S)=[\frac{1}{1+S^{2}L_{r}C_{r}}][\frac{1-d}{S^{2}L_{c}C_{c}+\frac{SL_{c}}{R_{c}}+(1-d)^{2}}]
\end{equation}
From (20), the steady state charging system transfer function yields to
\begin{equation}
G_{ss}=G(0)=\frac{1}{1-d}
\end{equation}
(21) is similar to the dc voltage gain of a generic boost converter.
\section{Equivalent Switching Converter Model of the Proposed Inverter Circuit}
\begin{figure}[!t]
	\centering
	\includegraphics[height=1.3in ,width=3.5 in]{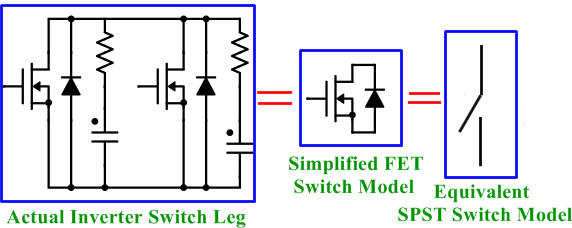}
	\caption{Simplified model of the switch used in the proposed power inverter circuit}
	\label{fig8}
\end{figure}
\begin{figure}[!t]
	\centering
	\includegraphics[height=5in ,width=3.2 in]{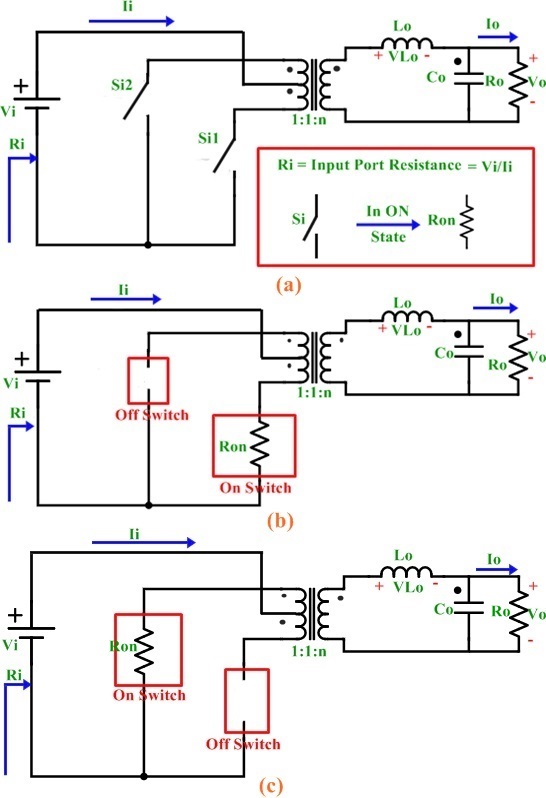}
	\caption{Switching converter model of the power inverter circuit; (a) switching converter model, (b) equivalent model during subinterval-1 and (c) equivalent model during subinterval-2}
	\label{fig9}
\end{figure}

This section subsumes derivation and analysis of an equivalent switching converter model of the proposed 4-switch push-pull inverter circuit. There are two switch legs- leg1: $Q_{i1}-Q_{i2}$ and leg2: $Q_{i3}-Q_{i4}$ in the inverter circuit. For effective power transformation, $Q_{i1}$ and $Q_{i2}$ are switched-on simultaneously whereas $Q_{i3}$ and $Q_{i4}$ are switched-on simultaneously. Inverter switch leg1 and leg2 are controlled by two alternating PWM pulses. Each of inverter 4-switches is a MOSFET with an anti-parallel diode constituting a current-bidirectional switch. Fig. \ref{fig8} presents a simplified model of the inverter switch in which a leg consisting of two switches is represented by a single pole single throw (SPST) switch model. Considering device losses, a closed switch is represented by an on-state resistance, $R_{on}$. Fig. \ref{fig9} shows the equivalent switching converter model of the push-pull inverter circuit. Here $S_{i1}$ presents the switch leg: $Q_{i3}$-$Q_{i4}$ and $S_{i2}$ presents the switch leg: $Q_{i1}$-$Q_{i2}$. During subinterval-1 or $dT_{s}$ interval, where d is the duty ratio and $T_{s}$ is the switching period, $S_{i1}$ is on (closed switch) and during subinterval-2 or $(1-d)T_{s}$ interval, $S_{i2}$ is on (closed switch). The circuit modelling approach assumes small ripple approximation. During $dT_{s}$ interval, the average inductor voltage is
\begin{equation}
V_{Lo}=nV_{i}-nI_{i}R_{on}-V_{o}
\end{equation}
 During $(1-d)T_{s}$ interval, the average inductor voltage is
\begin{equation}
	V_{Lo}=-nV_{i}+nI_{i}R_{on}-V_{o}
\end{equation}
According to inductor volt-second balance theory,
\begin{subequations}
	\begin{align}
	0=[nV_{i}-nI_{i}R_{on}-V_{o}][d]+[-nV_{i}+nI_{i}R_{on}-V_{o}][1-d]\\
	V_{o}=nd[V_{i}-I_{i}R_{on}]
	\end{align}
\end{subequations}
Hence the voltage gain of the circuit is
\begin{equation}
G_{v}=\frac{V_{o}}{V_{i}}=nd[1-R_{on}\frac{I_{i}}{V_{i}}]=nd[1-\frac{R_{on}}{R_{i}}]
\end{equation}
Here input port resistance, $R_{i}=\frac{V_{i}}{I_{i}}$. In case of lossless ($R_{on}\approx0$) switching devices,
\begin{equation}
G_{v}=\frac{V_{o}}{V_{i}}=nd
\end{equation}
By definition the power efficiency of the inverter is
\begin{subequations}
	\begin{align}
	\eta_{i}=\frac{P_{o}}{P_{i}}\times 100 \%=\frac{\frac{V_{o}^{2}}{R_{o}}}{V_{i}R_{i}}\times 100 \%\\
	\eta_{i}=\frac{n^{2}d^{2}[V_{i}-I_{i}R_{on}]^{2}}{R_{o}V_{i}I_{i}}\times 100 \%
	\end{align}
\end{subequations}
Here $P_{i}$ and $P_{o}$ are the input and output power of the circuit respectively. In case of lossless ($R_{on}\approx0$) switching devices, 
\begin{equation}
\eta_{i}=\frac{n^{2}d^{2}}{R_{o}}\frac{V_{i}}{I_{i}}\times 100 \%=n^{2}d^{2}\frac{R_{i}}{R_{o}}\times 100 \%
\end{equation}
From (25) and (28) it can be concluded that
\begin{equation}
G_{v}, \eta_{i}=f(n,d)
\end{equation}
(29) shows that inverter voltage gain and efficiency are functions of the transformer turns ratio and switching signal duty ratio.

\section{Simulation Results}
	 \begin{figure}[b!]
	 	\centering
	 	\includegraphics[height=3in ,width=3.5 in]{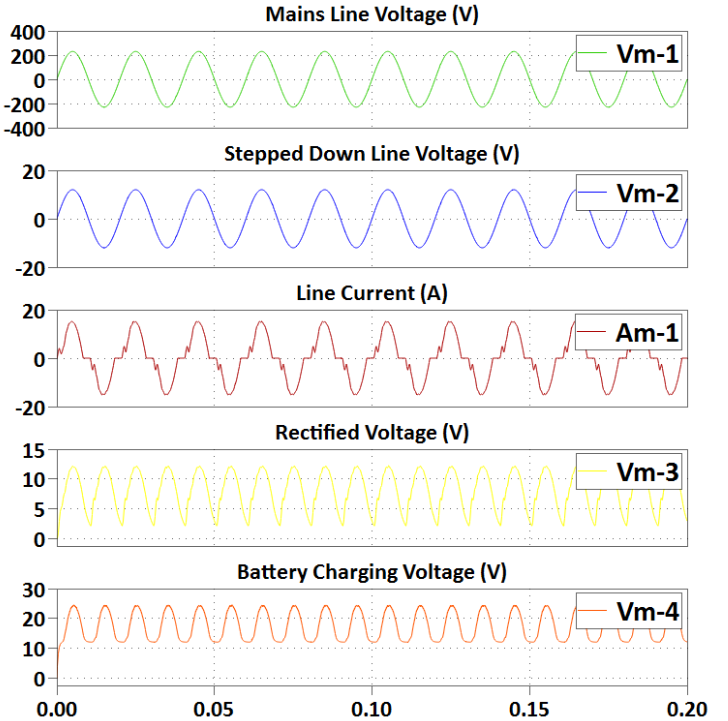}
	 	\caption{Simulated waveforms of the mains line quantities, rectified voltage and battery charging voltage}
	 	\label{fig10}
	 \end{figure}
	 \begin{figure}
	 	\centering
	 	\includegraphics[height=3in ,width=3.5 in]{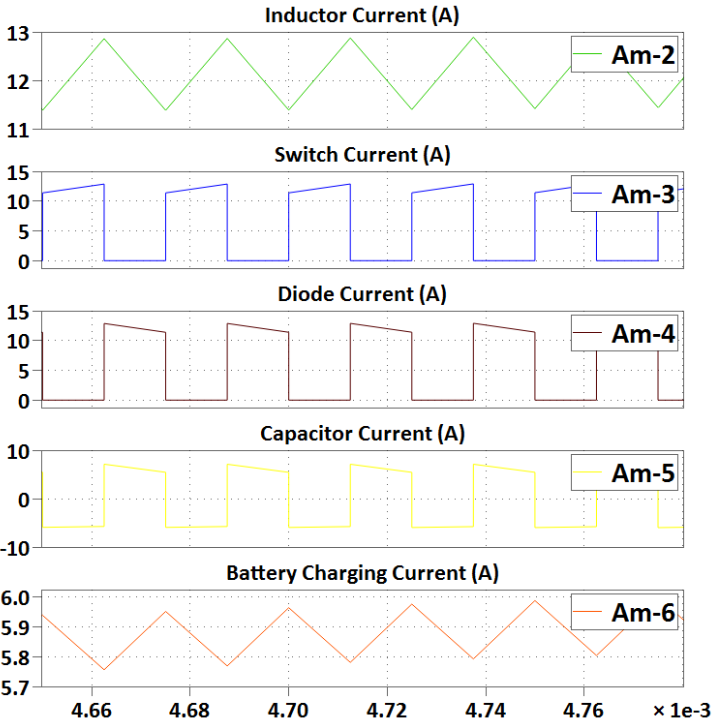}
	 	\caption{Simulated current waveforms across the inductor, switch, diode, capacitor and output port of the boost converter}
	 	\label{fig11}
	 \end{figure}
	 
	 \begin{figure}
	 	\centering
	 	\includegraphics[height=3in ,width=3.5 in]{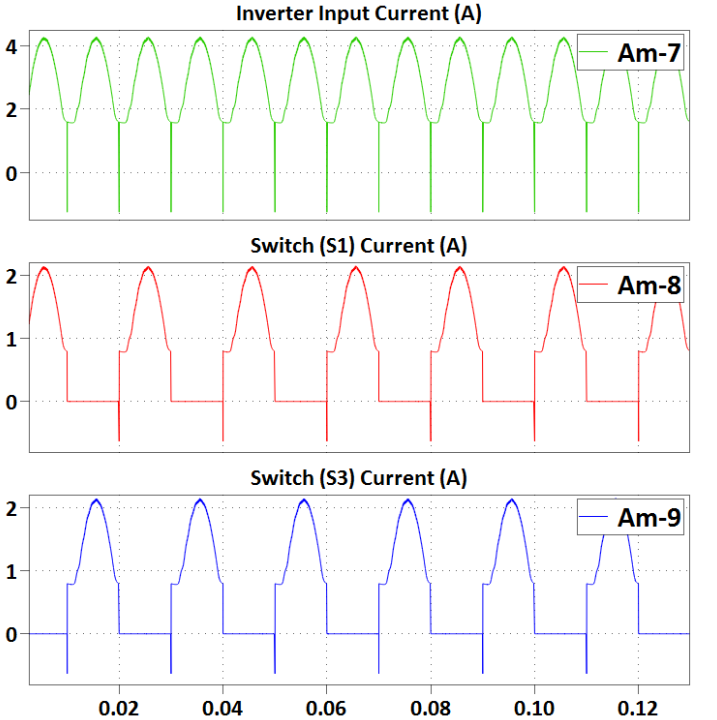}
	 	\caption{Simulated waveforms of inverter input current, current through switch $S1$ ($Q_{i3}$) and current through switch $S3$ ($Q_{i1}$) for an R-L load of R = 500 $\Omega$ and L = 27 mH}
	 	\label{fig12}
	 \end{figure}
	 
	 \begin{figure}
	 	\centering
	 	\includegraphics[height=3in ,width=3.5 in]{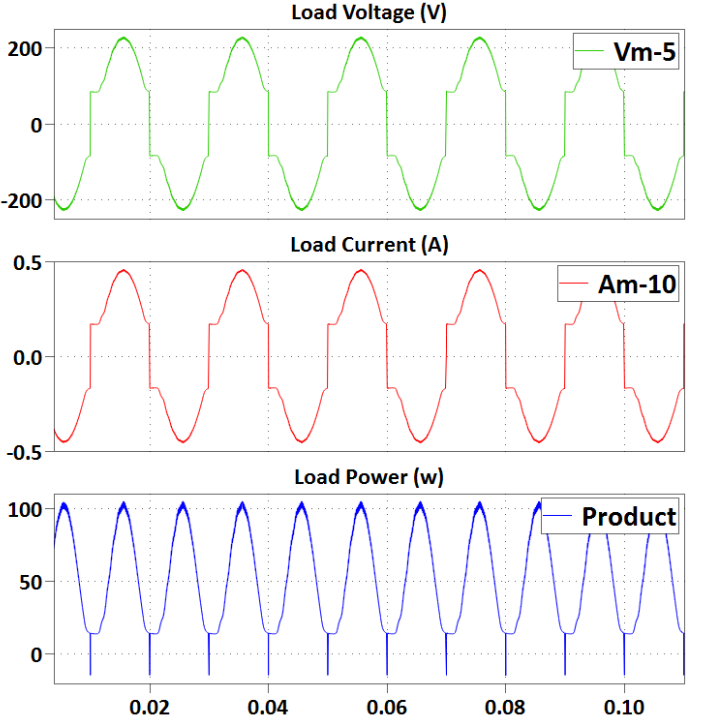}
	 	\caption{Simulated waveforms of the load voltage, load current and consumed power of the inverter for an R-L load of R = 500 $\Omega$ and L = 27 mH}
	 	\label{fig13}
	 \end{figure}
	 \begin{figure}
	 	\centering
	 	\includegraphics[height=3in,width=3.5 in]{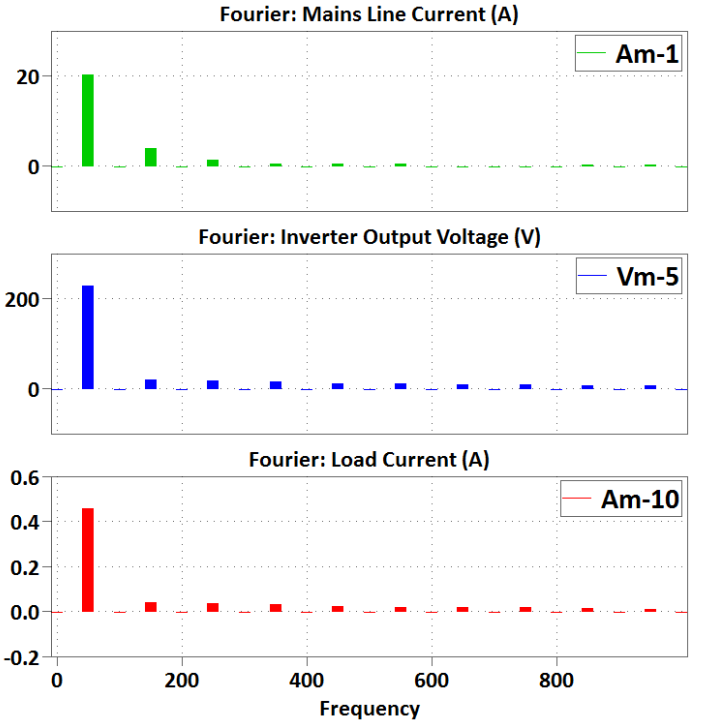}
	 	\caption{Simulated Fourier spectra of the mains current, inverter output voltage and output current for a load of R = 500 $\Omega$ and L = 27 mH}
	 	\label{fig14}
	 \end{figure}	 
The proposed 4-switch push-pull inverter based emergency back-up power supply with grid-interfaced battery charging circuit has been simulated in PLECS for an R-L load of R = 500 $\Omega$ and L = 27 mH with a maximum power consumption of nearly 104 w. In the design as presented in Fig. \ref{fig1}, the specifications of the circuit components selected for simulation are - $T_{x1}$ = 230 V - 12 V, $L_{r}$ = 0.1 mH, $C_{r}$ = 500 $\mu$F, $L_{c}$ = 0.95 mH, $C_{c}$ = 47 $\mu$F, $Q_{c}$, $Q_{i1}-Q_{i4}$ \& $Q_{r}$ = n-channel enhancement type MOSFETs, $R_{s1}-R_{s4}$ = 225 $\Omega$, $C_{s1}-C_{s4}$ \& $C_{sr}$ = 10 nF, $T_{x2}$ = 12 V - 230 V, $L_{o}$ = 21.2 mH, $C_{o}$ = 470 $\mu$F, $R_{r1}$ = 1 k$\Omega$, $R_{r2}$ = 12 k$\Omega$ and $V_{dd}$ = 12 V. The battery is a 12 V voltage source and the switching relays are the relay models available in PLECS library. However, the switching PWM signals have been generated independently by symmetrical PWM blocks of PLECS. Figs. \ref{fig10} - \ref{fig14} present the simulated waveforms of the system components. However, in Figs. \ref{fig10} - \ref{fig13} the x-axis represents the simulation time in s whereas in Fig. \ref{fig14} the x-axis contains the frequency values in Hz. 

\begin{table*}[!t]
	\centering
	\caption{Simulated Data of Harmonic Analysis}
	\begin{tabular}{p{0.5in}|p{1.2in}|p{1.2in}|p{1.2in}}
		\hline
		Harmonic Order, $n$ & Mains Harmonic Current Content, $I_{n,m}$(A) & Inverter Output Harmonic Voltage Content, $V_{n,l}$(V) & Load Harmonic Current Content, $I_{n,l}$(A) \\
		\hline
		1 & 20.2184 & 228.361 & 0.45473\\
		\hline
		3 & 3.97007 & 19.3039 & 0.0385546\\
		\hline
		5 & 1.25227 & 17.5455 & 0.0349619\\
		\hline
		7 & 0.367276 & 14.1336 & 0.0280666\\
		\hline
		9 & 0.476162 & 11.1722 & 0.0220854\\
		\hline
		11 & 0.337621 & 9.68919 & 0.0190475\\
		\hline
		13 & 0.140560 & 8.78997 & 0.0171669\\
		\hline
		even & 0 & 0 & 0
	\end{tabular}%
		\label{table2}%
	\end{table*}%
	\begin{table*}[!t]
		\centering
		\caption{Simulated Data of the Performance Parameters for a Maximum Load Power Consumption of 103.59 w}
		\begin{tabular}{p{1in}|p{1.5in}|p{1.5in}|p{1.2in}}
			\hline
			Sending-End Power Factor, $ PF_{s}$ & Mains Current THD, $I_{mTHD}$ (\%) & Inverter Output Voltage THD, $V_{lTHD}$ (\%) & Load Current THD, $I_{lTHD}$ (\%) \\
			\hline
			0.9639 & 20.8818 & 15.0183 & 14.9827
		\end{tabular}%
		\label{table3}%
	\end{table*}%

Table \ref{table2} presents the simulation outcomes of harmonic analysis of mains line current, inverter output voltage and load current. Table \ref{table3} presents the performance evaluation of the proposed system in terms of sending-end power factor, mains current THD, inverter output voltage THD and load current THD. By definition, sending-end power factor is
\begin{equation}
PF_{s}=\frac{P_{R}}{S_{A}}=\frac{V_{s}I_{s}cos\phi}{V_{s}I_{s}}=cos\phi
\end{equation}
Here $P_{R}$ is the real power (w), $S_{A}$ is the apparent power (VA), $\phi$ is the angle between voltage, $V_{s}$ and current, $I_{s}$. By definition, current THD (\%) and voltage THD (\%) are     
\begin{equation}
I_{THD}=\frac{\sqrt{I_{2}^2+I_{3}^2+I_{4}^2+I_{5}^2+I_{6}^2+...}}{I_{1}} \times 100 \%
\end{equation}
\begin{equation}
V_{THD}=\frac{\sqrt{V_{2}^2+V_{3}^2+V_{4}^2+V_{5}^2+V_{6}^2+...}}{V_{1}} \times 100 \%
\end{equation}
Any power quantity can be represented by $H$ (= $V$ and $I$). Here $H_{1}$ is the fundamental or base frequency component, $H_{2}$, $H_{3}$, $H_{4}$, $H_{5}$ etc are the 2nd, 3rd, 4th, 5th etc order harmonic components respectively. Table \ref{table3} shows that in simulation, the sending-end power factor and mains current THD of the system are around 0.96 and 21 \% respectively which are considerably appropriate for a rectifier-link dc boost converter with PWM switching control application. In simulation, for an R-L load with an approximated maximum power consumption of 104 w, the load voltage THD and load current THD values are about 15 \% which are considerably low and acceptable for an efficient power supply system.

\section{Conclusion}
In this paper design of a grid-integrated emergency back-up power supply for medium power utility applications has been articulated. A 4-switch push-pull inverter circuit topology has been designed and simulated which generates modified square wave voltage signals at grid frequency. The battery is charged by a rectifier-fed PWM boost derived charging circuit. The instantaneous transfer switching from the grid to the customized power supply system has been substantiated by a changeover relay. An SPDT relay switching circuit forms a battery charge controller to provide over-voltage and over-charge protection. The switching converters have been controlled by PWM signals. Respective state space and switching converter models of the charging circuit and power inverter have been derived. The proposed design has been simulated in PLECS. Laboratory evaluation and a more compact and high power sine wave generating power system with feedback control feature are future scopes of this work. 

\ifCLASSOPTIONcaptionsoff
  \newpage
\fi

\end{document}